# Hydrogenation Dynamics of Twisted Carbon Nanotubes

*Jose M. de Sousa[‡], Pedro A. S. Autreto[‡], Douglas S. Galvao[*]*

Applied Physics Department, State University of Campinas, Campinas/SP, 13083-970, Brazil.

**Keywords:** carbon nanotubes, reactive molecular dynamics, fracture

Carbon Nanotubes (CNTs) are one of the most important materials in nanotechnology. In some of their technological applications (electromechanical oscillators and mechanical actuators for artificial muscles, for instance), it is necessary to subject them to large deformations. Although this frequently happens in air, there are only few studies about the interaction of deformed CNTs with the atmosphere and the dynamics of these processes has not yet been addressed. In this work, we have investigated, through fully atomistic reactive molecular dynamics simulations, the process of hydrogenation of highly twisted CNTs. Our results show that hydrogenation effective ratio is directly related to the tube twist angle values and can lead to twisted tube fractures with well defined patterns (unzip-like). Our results also show that these fracture processes can be exploited to controllably produce graphene nanoribbons.

---

[*] Telephone: +5519-35215373 - Email:galvao@ifi.unicamp.br

**INTRODUCTION**

Carbon-based nanostructures are among the most important materials in nanoscience[1–4]. The carbon hybridization versatility allows the existence of many allotropic forms, such as; diamond[5], fullerenes[6], carbon nanotubes (CNTs)[7], graphynes[8–10] and graphene[11]. These materials exhibit unique mechanical[12–15], chemical[15-17] and, thermal[18–20] properties, which have been exploited in a large variety of technological applications[21,22].

This is especially true for CNTs[23-27]. Their excellent electrical[28,29] and mechanical properties (some elastic properties two orders or magnitude higher than metallic materials[30]) have been used to build a variety of applications including sensors[31], actuators[32,33], springs[34], among others. CNT chiralities[35], chemical functionalizations[36,37], and mechanical deformations, such as; bending and torsions[38–40], can be used to tune some of the CNT properties.

CNTs are very stable structures[41], although their most reactive parts (ends, the "caps"[42]) can be etched through oxidation processes[41,43,44]. Ruoff and co-workers[45] showed, from Functional Density Theory (DFT) calculations, that CNT reactivity is directly related to tube curvatures. They suggested that reactivity could be enhanced through structural deformations that could increase local curvature and/or strain, as for instance tube twisting[45]. In spite of the large number of CNT studies, the detailed dynamics of these processes has not yet been investigated.

This kind of study is important because it is possible today to controllably subject CNTs to different types of mechanical deformations, as already experimentally demonstrated for example in electromechanical oscillators[46,47] and mechanical actuators for artificial muscles[2,48–50]. Understanding how CNT physical and chemical (especially reactivity with the environment) properties change under mechanical deformations is of fundamental importance in order to design better and more reliable CNT-based devices. It is also important to help to develop new approaches to manipulate and to tune CNT properties.

In this work we have investigated, through fully atomistic molecular dynamics (MD) simulations, the relative reactivity of twisted CNTs. In particular, we have investigated how the CNT hydrogenation dynamics (structural stability and fracture patterns) depend on the tube twisting angle values (θ angle, Figure 1).

Our results show that the level of effective hydrogenation (number of formed hydrogen-carbon bonds) is directly proportional to the CNT twist angle values and increases with temperature, up to the limit of structural degradation. Our results also show that the use of extensive hydrogenation on twisted CNT can be a novel way to produce graphene nanoribbons (GNRs) (large aspect ratio graphene strips obtained from unzipped/collapsed CNTs). There is a great interest in GNRs because of, in contrast to pristine graphene, they exhibit non-zero electronic gap (needed for digital electronic applications) and their electronic and spintronics properties can be easily tuned[51–55]. In fact, different approaches for GNRs production have been already developed, such as; chemical attacks[51–53], mechanical strain/stress[54] and, most recently, through high velocity impact collision[55].

## METHODOLOGY

In order to investigate the hydrogenation dynamics of twisted CNTs we carried out MD simulations using fully atomistic reactive molecular dynamics methods (ReaxFF)[56] as available in the open source code Large-scale atomic/ Molecular Massively Parallel Simulator (LAMMPS)[57] package. Developed by van Duin, Goddard III and co-workers, ReaxFF is a reactive force field. Reactive meaning that it allows the simulation of creating/breaking chemical bonds, and thus it can be used in problems involving chemical reactions (such as, in the hydrogenation processes we are investigating here). Similarly to other non-reactive force fields (for example, MM3 and MM4[58,59]), in ReaxFF the energy contributions are divided into separated terms, such as; valence angles, bonds stretching, van der Waals interactions, Coulomb, among others[56]. ReaxFF is parameterized using calculations based on Density Functional Theory (DFT) methods, and typically the difference between the heats of formation obtained from ReaxFF calculations and experimental results, for unconjugated and conjugated systems, is around 3.0 kcal/mol[56]. Due to its relative low (in comparison to DFT, for instance) computational cost, ReaxFF can be used to study large systems. It has been applied with success in several investigations of carbon-based materials at the atomic scale[54,55].

For all our MD simulations, we used a canonical ensemble (NVT), coupled to a Noosé-Hover thermostat[60]. Typical run time simulations were 2.5 ns, with time steps of 0.1 fs. Different temperatures were considered (from 0 up to 1000 K) in order to determine the temperature influence on the hydrogenation dynamics.

Our simulated systems consisted of CNTs (different chiralities, diameter and length values were considered), at different twisted states (we considered the θ twist angle values of: 0, 180, 360, 450, 720 degrees, respectively) immersed into a hydrogen gaseous atmosphere (Figure 1). In order to speed up our simulations atomic hydrogen atoms were used and their recombination (H-H) was not allowed.

**RESULTS AND DISCUSSION**

In Figure 2 we present representative MD snapshots of the obtained hydrogenated twisted carbon nanotubes for different twist angle values (0º, 180º, 360º, 450º, 720º), as a function of time simulation (see also video01 in the supplementary material). From the figure it is possible to notice that for the non-twisted (0 degree) case, only a very small number of bonded hydrogen atoms are observed. This number significantly increases with the twist angle values. In Figure 2 we can also notice that the hydrogen covering occurs mainly at the regions of extensive deformations (higher curvature), in agreement with Ruoff and co-authors predictions[45].

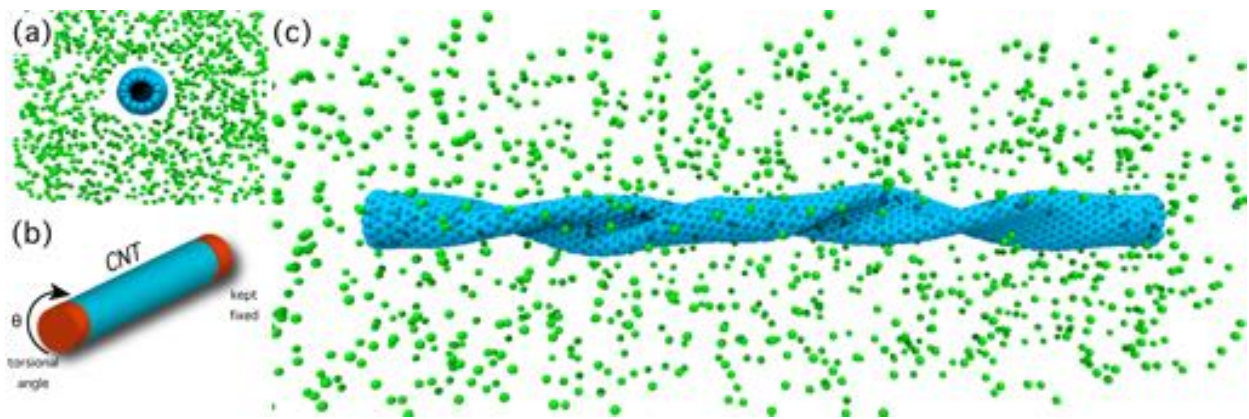

**Figure 1.** Simulated systems: twisted carbon nanotubes (CNTs) immersed into an atomic hydrogen (green atoms) atmosphere (a) and (c) (top and side views). In the inset (b), the twist angle definition. In order to apply the twist load the tube ends (indicated in red) were kept fixed during the hydrogenation MD simulations.

In this sense, highly twisted CNTs will adsorb more hydrogen, since topologically they will present (proportionally) a larger deformed area. This behavior is better evidenced in Figure 3, where we present the ratio of bonded hydrogen atoms per available carbon ones, as a function of the simulation time for different tube twist angle values. While for non-twisted CNTs the hydrogen bonding is not significant, for the case of 720º it can reach ~ 20%.

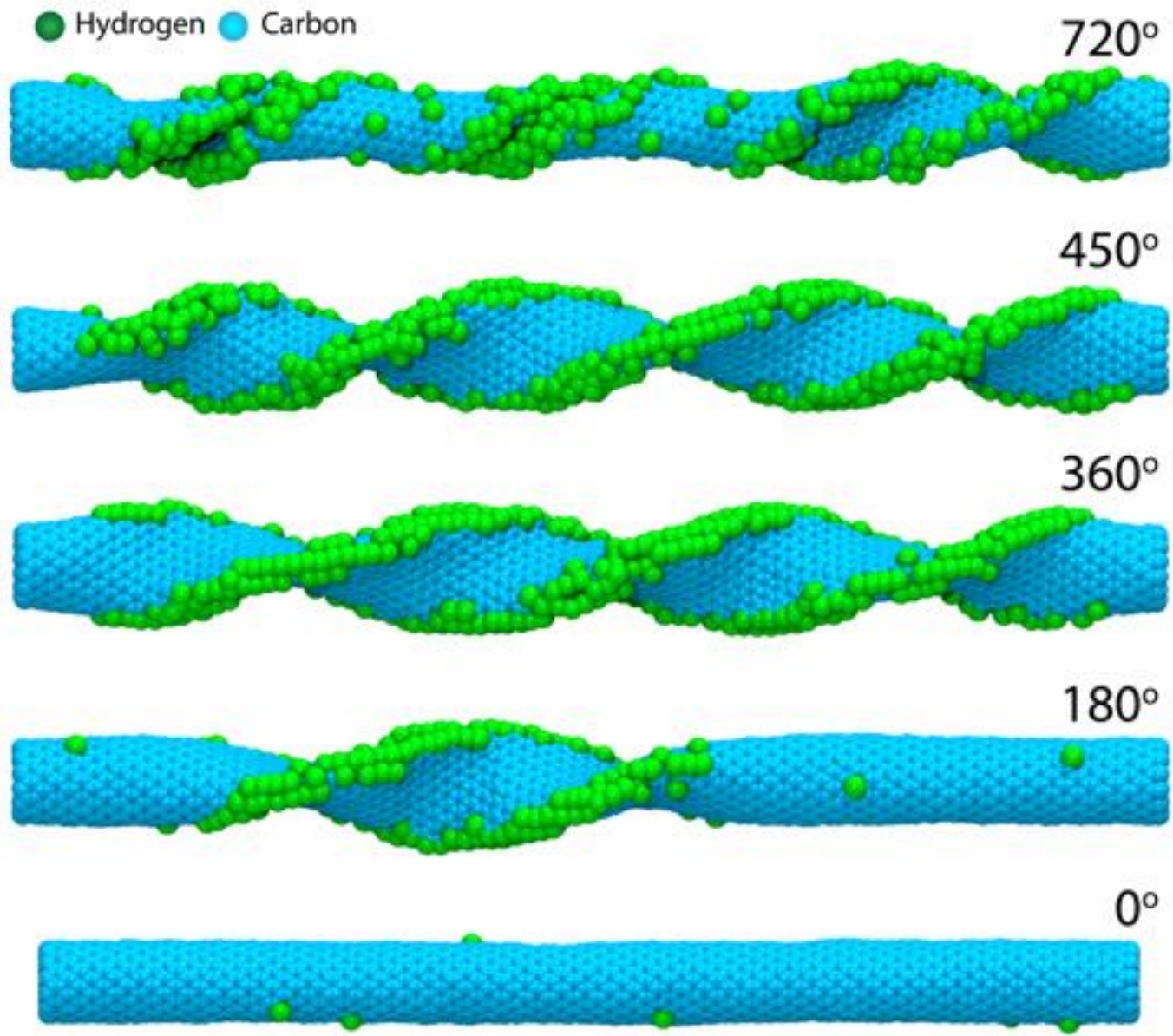

**Figure 2.** Representative MD snapshots showing the final stages of functionalized (hydrogenated) twisted carbon nanotube, for different twist angle values. Increasing the twist angle results in a significant increase in the number of bonded hydrogen atoms. Results from simulations at 600 K, zigzag CNT(14,0) with length and diameter values of ~ 190 Å and 90 Å, respectively.

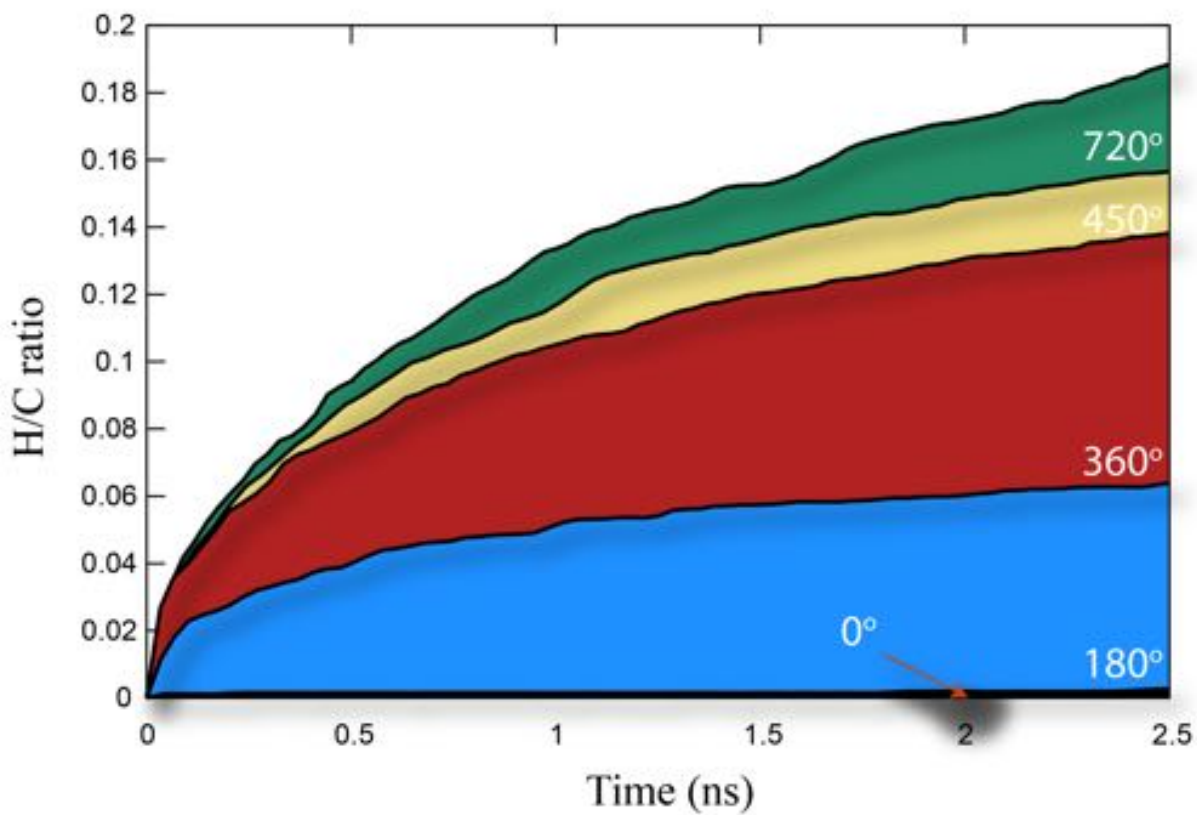

**Figure 3.** Ratio of bonded hydrogen atoms per available carbon ones, as function of simulation time, for different twist angle values.

The hydrogenation process of the deformed regions in the twisted CNTs also induces a process of C-C breaking bonds. In Figure 4 we present the percentage of C-C breaks as function of the simulation time, for different twist angle values. A schematic pathway of these bond breakings is presented at the top of Figure 4. We can see from Figure 4 that the H/C ratio continuously increases in time. Also, the general trends observed at the bottom part of Figure 4 are the same ones observed in Figure 3. This is a direct consequence that the bond breakings are intrinsically related to hydrogen adsorption in highly deformed areas. For the non-twisted CNTs the small amount of hydrogen bonds does not lead to tube fracture. On the other hand, for the twisted tubes the fractured areas (measured by the number of broken C-C bonds) increase proportional to the twist angle values, reaching 4% of broken C-C bonds for 720º. Random fractures commonly observed in functionalized CNTs were not observed in the non-hydrogenated areas. More importantly, the fracture occurs along the pathways of the hydrogenated areas, generating patterns very similar to the ones observed in the processes of unzipping CNTs[51,53,54]. Ruoff and co-authors[45] proposed that CNT areas with high curvature would be more reactive and that it is exactly what we observed here. The tube twisting creates well-defined (twisted lines) deformed areas with high curvature, which are the areas preferentially attacked by the hydrogen atoms and when the fractures occur. The fracture patterns follow these twisted lines, leading to tube strain driven unzipping (see also video02 and video03 in the supplementary material).

These results also suggest that the hydrogenation of twisted CNTs can be used to produce, in a controllable way, graphene nanoribbons (GNRs) with adjustable dimensions: edges (armchair or zigzag) and shapes. As discussed above the hydrogenation of the twisted areas lead to their fracture and tube structural collapse. When the twist load is released the collapsed tube tend to untwist forming bilayer-like GNRs, but depending on the level of incompletely fractured areas (see Figure 5), this is only partially achieved. Better formed GNRs can be obtained using high temperatures (see video03 in the supplementary material for the case of T= 1000 K) and/or longer hydrogenation time. The bilayer separation can be obtained using standard chemical processes[51-53].

The bond breaking process can be better understood from Figure 6 where we present the calculated von Mises stress values for twisted CNTs and their hydrogenated forms. The von Mises stress allows us to easily evaluate the level of local deformation and it is very helpful to understand fracture patterns[54]. From this Figure we can see that the hydrogen atoms preferentially attach to the twist lines as discussed before. This creates region of high stress (red regions in the Figure) that subsequently causes bond breaking (releasing stress) and generates fractured areas leading to GNR-like structures.

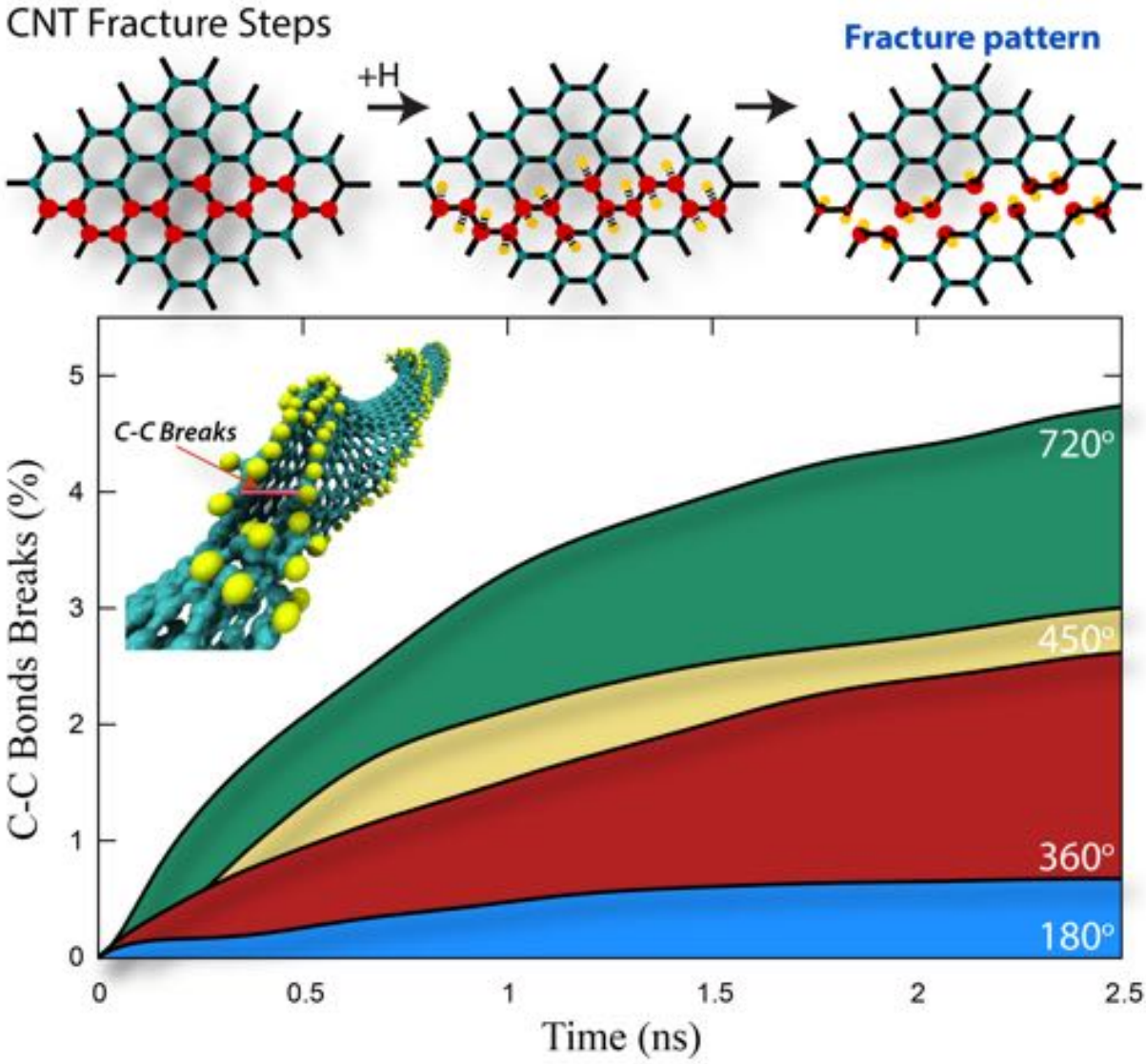

**Figure 4.** TOP: Scheme of the fracture pathway patterns induced by hydrogen bonding into twisted CNTs. BOTTOM: Percentage of broken bonds as function of the time simulation for different twist angle values.

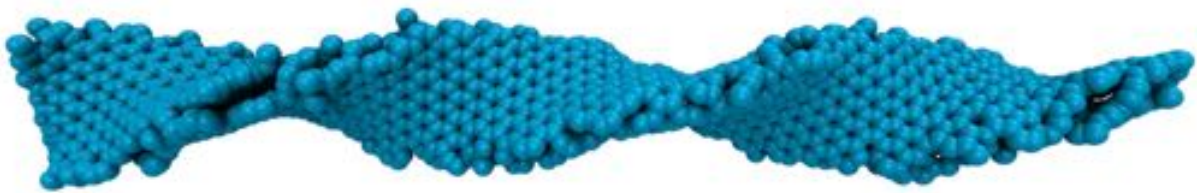

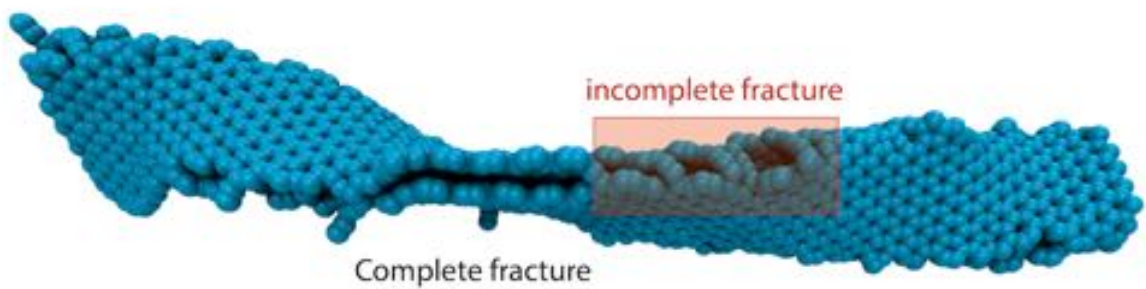

**Figure 5.** Representative MD snapshots showing the initial and final stage of the hydrogenation process, for the case of 450°. For clarity the hydrogen atoms were made transparent. After the twist load release (final state), the tube tends to untwist, but fractured and incomplete fractured areas coexist.

**CONCLUSIONS**

We have investigated, through fully atomistic molecular dynamics simulations, the hydrogen dynamics of highly twisted CNTs. Our results show that the hydrogen bond formation (chemical reactivity) increases proportional to the twist angle and temperature values. These hydrogen bonds form preferentially on the regions of high curvature (twist lines). Also, this increases the local strain accumulates stress in some parts of the tube wall which will be relieved C-C bond breaks, leading to the existence of fractured and partially fracture regions forming bilayer graphene nanoribbon-like structures. These general conclusions are not chirality, diameter and tube length dependent. Our results suggest that extensive hydrogenation of highly twisted CNTs can be a novel way to controllably produce graphene nanoribbons from single-walled CNTs, which has been proved to be very difficult from available chemical and physical methods[51-55].

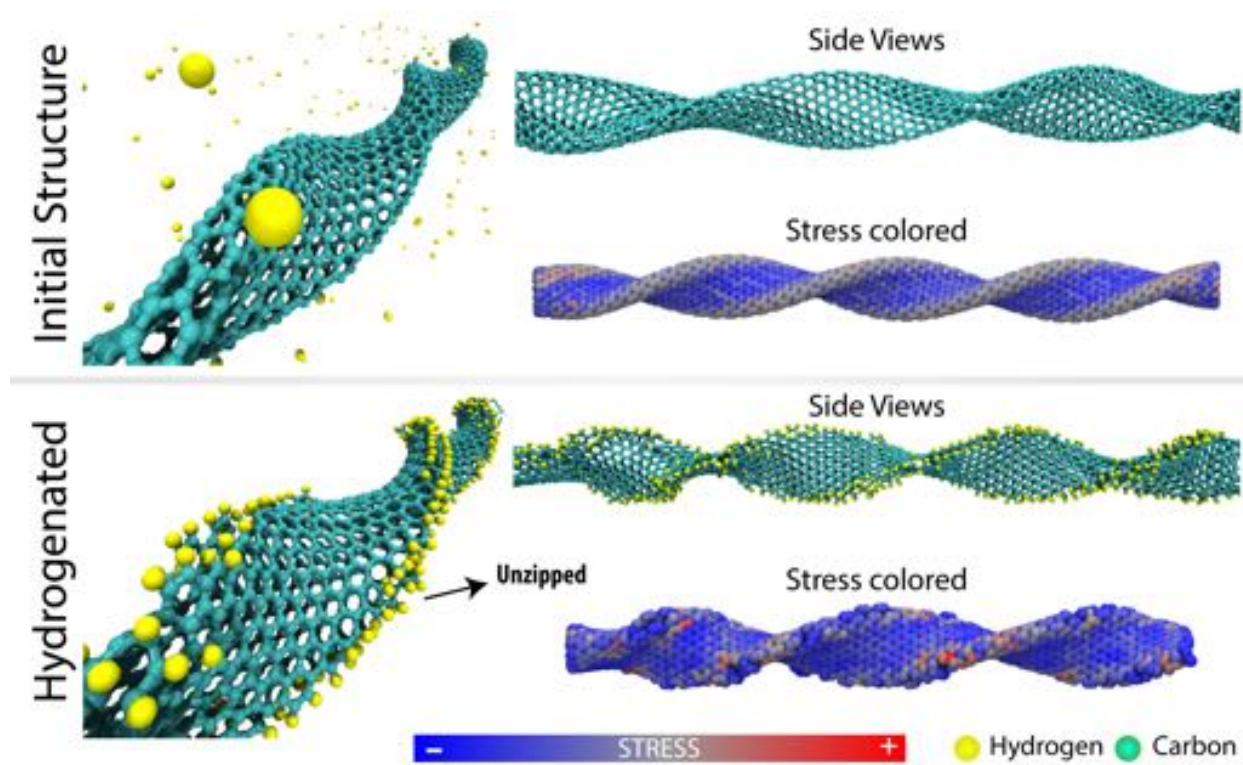

**Figure 6.** Snapshots showing the stress von Misses in twisted CNTs. (a) Initial conditions without hydrogenation where the gray color stress shows the most stressed region. (b) Hydrogenated structure at the end *2.5ns*.

**AUTHOR INFORMATION**
**Corresponding Author**
*Email:galvao@ifi.unicamp.br

**Author Contributions**

‡These authors contributed equally for this work. J. M. de Sousa and P.A.S. Autreto carried out simulations and analyzed the data. All authors were involved in discussing the results and writing the paper.

ACKNOWLEDGMENT

This work was supported in part by the Brazilian Agencies CAPES, CNPq and FAPESP. The authors thank the Center for Computational Engineering and Sciences at Unicamp for financial support through the FAPESP/CEPID Grant # 2013/08293-7.